\documentstyle[prd,aps,floats,epsf]{revtex}

\def\be{\begin{equation}}
\def\ee{\end{equation}}
\def\bea{\begin{eqnarray}}
\def\eea{\end{eqnarray}}

\begin{document}
\preprint{astro-ph/yymmddd}
\draft 

%
%
 
%
\renewcommand{\topfraction}{0.99}
\renewcommand{\bottomfraction}{0.99}
\twocolumn[\hsize\textwidth\columnwidth\hsize\csname 
@twocolumnfalse\endcsname
  
\title
{\Large Cosmological Effects of a Class of Fluid Dark Energy Models}
       
\author{Daniela Carturan$^{1,2}$ and Fabio Finelli$^{2}$} 
\address{~\\$^1$Department of Astronomy, University of Bologna, Via
Ranzani 1, I-40127 Bologna, Italy;
~\\$^2$IASF-CNR/Bologna, Via Gobetti 101, I-40129, Bologna, Italy}
\date{\today} 
\maketitle
\begin{abstract} 
We study the impact of a generalized Chaplygin gas as a candidate for
dark energy on CMB anisotropies.
The generalized Chaplygin gas 
is a fluid component with an exotic equation of state, 
$p = - A/\rho^\alpha$ (a polytropic gas with negative constant
and exponent). Such component interpolates in time between 
dust and a cosmological constant, with an intermediate behaviour as 
$p=\alpha \rho$. Perturbations of this fluid are stable on small scales,
but behave in a very different way with respect to standard quintessence.
Moreover, a generalized Chaplygin gas could also
represent an archetypal example of a phenomenological unified models of
dark energy and dark matter. The results presented here show how the 
CMB anisotropies induced by this class of models differ from a
$\Lambda$CDM model.
\end{abstract}

]

\vskip 0.4cm
{\bf 1.} According to the present recipe to explain observational data,
two dark components seem to fill the universe up to $95 \%$ of the total
content. The baryon density is indeed very small, $\Omega_b h^2 \sim 0.02$
\cite{baryon}, assuming a flat Universe. 
This situation has been slowly reached over decades.
In addition to cold dark matter (CDM), 
in the $90$s a cosmological constant term $\Lambda$ was called into
play to explain the recent accelleration of the universe 
indicated by the Supernova data \cite{SN}, but then confirmed by other
observations. 

Avoiding to explain on theoretical grounds the embarassing smallness of a
cosmological constant $\Lambda$ constrained by observations, a scalar
field $Q$ \cite{rp}, dubbed quintessence, was suggested in order to
explain the
accellerating Universe \cite{accellera}. Because of the different
background evolution and of the presence of fluctuations, the challenge is
to distinguish $\Lambda$ from quintessence, for instance through the
Cosmic Microwave Background (CMB) anisotropies \cite{lversusq}. 

An alternative to quintessence for modelling a dark energy component could
be a perfect fluid (henceforth PF) with a generic pressure $p=p (\rho)$
(not being linear in energy density $\rho$ as for barotropic fluids)
whose energy momentum tensor is:
\be
T_{\mu \nu} = p g_{\mu \nu} + (\rho + p) u_\mu u_\nu
\ee
where $g_{\mu \nu}$ is the metric and $u_\mu$ is the fluid velocity
($u_\mu u^\mu = -1$).
From a theoretical point of view a scalar field description would be
preferrable in the logic to relate an accelerating universe with a
fundamental quantum. From a phenomenological point of view the reasons to
prefer one over the other are less obvious. An exotic fluid
capable to develop a negative pressure at late times may represent as well
the effective degree of freedom which drives the present accelleration of
the universe.
In particular, a PF model with this property would allow to explore the
possibility that dark energy clusters on small scales. Indeed, it is
important to note that, by parametrizing dark energy with an uncoupled
scalar field with ordinary kinetic term, one does implicitly assume no
clustering of dark energy on scales smaller than the Hubble radius.

Among the class of PF models which could work as a dark energy component -
in principle one could design suitable pressure profiles $p(\rho)$
instead of potentials $V(\phi)$ for a quintessence field $\phi$ - 
the Chaplygin gas \cite{chaplygin} has recently received a lot of
attention \cite{kmp}. 
A Chaplygin gas is characterized by a pressure $p_X$ related to the
energy density $\rho_X$ in the following way:
\be
p_X = - \frac{A}{\rho_X^\alpha}
\label{pressure}
\ee
with $\alpha = 1$. Even if such exotic fluid was proposed in the context
of aereodynamics
\cite{chaplygin}, there are interesting connections with particle physics 
and d-branes \cite{jackiw}.
A Chaplygin gas is also equivalent to a tachyon field with a constant
potential \cite{frolov} and, at the homogeneous level, to a complex scalar
field \cite{tf} or to a quantum scalar field \cite{fvv} in the bottom
of a potential (called Thomas-Fermi approximation
in \cite{tf}). In this paper, we study a generalized version of
the Chaplygin gas (henceforth GCG) \cite{gcg} by considering $0 < \alpha
\le 1$ in Eq. (\ref{pressure}).

\begin{figure}
\epsfxsize=2.9 in \epsfbox{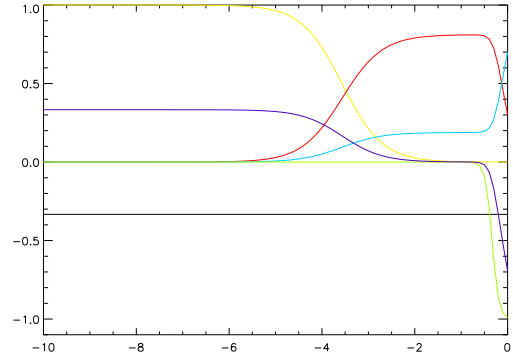}
\caption{Evolution of the background quantities $\Omega_r$ (yellow),
$\Omega_m$ (red), $\Omega_X$ (blue), $w_{\rm tot}$ (purple) and $w_X$     
(green) for a flat universe versus the scale factor. The horizontal black
line is $-1/3$, which denotes the threshold
for $w$ below which the universe expands accellerating. The parameters   
are $h=.7$, $\Omega_{r 0}=10^{-4}$, $\Omega_{m 0} \simeq .3$,
$\Omega_{X 0}=.7$, $\alpha=1$, $B/A = .01$.}
\label{fig1}
\end{figure}

{\bf 2.} In a Robertson-Walker metric
\be
ds^2 = - dt^2 + a(t)^2 ( \frac{dr^2}{1 - K r^2} + r^2 d \Theta^2  ) \,,
\ee
where $K = 0, \pm 1$ is the curvature of the spatial sections and
$\Theta$ is the solid angle, the energy conservation equation for a GCG
\be
\dot \rho_X + 3 H (\rho_X + p_X) = 0
\label{encons}
\ee
can be immediately integrated \cite{kmp,gcg}:
\be
\rho_X = \left( A + \frac{B}{a^{3 (1+\alpha)}}
\right)^{\frac{1}{1+\alpha}}
\,.
\label{solution}
\ee
where $A, B$ are constants with dimensions $[ M^{4 (1+ \alpha)} ]$.
We note that a GCG reduces to a $\Lambda$CDM model for $\alpha=0$
and to a sCDM for $A=0$.
We see that this fluid with an exotic equation of state
behaves like dust for small $a$ (when $B/A >> a^{3 (1+\alpha)}$, assuming 
$a=1$ at present time) and a
cosmological constant given by $A^{\frac{1}{1+\alpha}}$ in the opposite
limit ($B/A << a^{3 (1+\alpha)}$). By Taylor expanding in this limit
\cite{kmp,gcg} we obtain from Eqs. (\ref{solution}) and (\ref{pressure}): 
\begin{eqnarray}
\rho_X &\simeq& A^{\frac{1}{1+\alpha}} \left( 1 + \frac{B}{(1+\alpha) A 
\, a^{3(1+\alpha)}} + {\cal O} \left( \frac{B^2}{A^2} \right) \right)
\nonumber \\
p_X &\simeq& A^{\frac{1}{1+\alpha}} \left( - 1 + \frac{\alpha
B}{(1+\alpha) A \, a^{3(1+\alpha)}} + {\cal O} \left( \frac{B^2}{A^2}
\right) \right) \,.
\label{limit}
\end{eqnarray}
Therefore, in this limit a GCG behaves as a sum of a cosmological constant
and a perfect fluid characterized by $p = \alpha \rho$, sheding light upon
the physical meaning of the parameter $\alpha$. 
We note that the solution (\ref{solution}) to Eq. (\ref{encons}) is valid
for {\em any} $\alpha > -1$, but also for $\alpha < -1$ (a standard
polytropic gas). In this latter 
case the behaviour of such a PF interpolates between a cosmological
constant and dust. For $\alpha = -1$
Eq. (\ref{pressure}) describes the usual barotropic perfect fluid.
According to Eq. (\ref{pressure}), the equation of state $w_X$ defined as:
\be 
w_X = \frac{p_X}{\rho_X} = - \frac{A}{\rho_X^{1+\alpha}}   
\ee
decreases from the value $0$ to $-1$. An example of a background evolution
for a GCG as dark energy is given in Fig. (\ref{fig1}).

The presence of a GCG as a dark energy component could be distinguishable 
from a quintessence component because of the parametric form of the
pressure $p = p(\rho)$. The time derivative of the equation of state of
the dark energy component is important in the program of reconstructing
the total equation of state \cite{statefinder}.
Indeed, the time derivative of $w_X$ for a GCG is
\be
\dot w_X = 3 H (\alpha+1) w_X (1+w_X) 
\label{wxderiv}
\ee
while for a scalar field $\phi$ with potential $V = V(\phi)$ is:
\be
\dot w_\phi = 3 H (1 + w_\phi) (w_\phi - 1) - 2 \frac{\dot V}{\rho_\phi}
\,.
\ee
SN Ia observations can constraint the GCG as a candidate for dark energy,
as recently studied by different authors \cite{SNconstraint}. The purpose
of this letter is to show how CMB data could be more selective.

{\bf 3.} The behaviour of perturbations is a very interesting aspect
of the model.
When trying to build a model for an accelerating universe with a 
barotropic (constant equation of state $w$) perfect fluid,
one runs in the problem of
instabilities on short scales because of a negative sound speed for
perturbations. Infact, the sound speed is equal to the equation of
state $w$ and this must be negative ($w < - 1/3$) to explain the
accelleration.
This is the usual problem for a fluid description of domain walls and
cosmic strings. Quintessence models with scalar fields with a standard
kinetic term do not have this problem. The sound speed for a scalar field
is equal to $1$. K-essence models \cite{kessence} based on scalar
fields with a non-standard kinetic terms are different in this respect
\cite{garrmukha}, but still have positive sound speed. For a GCG the sound
speed $c_X^2$ for perturbations is 
\be
c_X^2 = \frac{\partial p}{\partial \rho} = \alpha
\frac{A}{\rho_X^{1+\alpha}} =
- \alpha w_X \,.
\label{soundx}
\ee
Therefore, because of its non-barotropic nature, perturbations of
a GCG are stable on small
scales even in an accellerating phase, and behave similarly to dust
perturbations when the gas is
in the dust regime. When the behaviour of the background Chaplygin gas is
of $\Lambda$-type, the sound-speed is $\alpha$. In order to avoid
causality issues, we shall consider $\alpha \le 1$. 
This latter constraint, with the requirement of positive sound speed
($\alpha > 0$), marks the interesting physical range of $\alpha$. 
We note that the possibility of having a non negative sound speed and
a negative equation of state, as happens for a GCG,
could open a new development in modelling a domain wall or cosmic string
network which is not plagued by short scale instability for
perturbations.

\begin{figure}
\epsfxsize=2.9 in \epsfbox{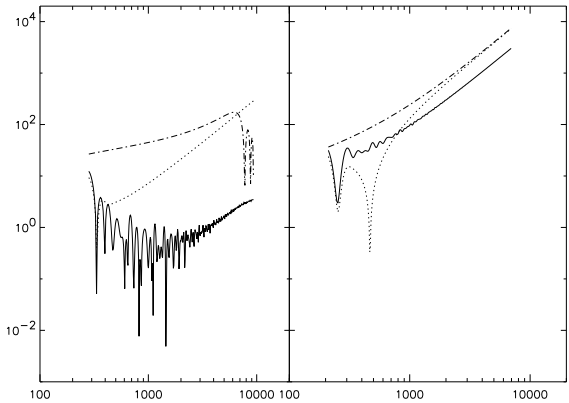}
\vspace{.3cm}
\caption{Density contrasts of photons (solid line), baryons (dashed
line), GCG (short dashed, on the left) and CDM (short dashed, on the
right) in a GCG model (left) and a $\Lambda$CDM model (right).}
\label{figcontrasti}
\end{figure}

The equations for the energy density contrast $\delta_X = \delta \rho_X 
/ \rho_X$ and the velocity potential $\theta_X$ in the
synchronous gauge are, according to Ref. \cite{mabert} and by using Eqs.
(\ref{wxderiv},\ref{soundx})
\begin{eqnarray}
\delta_X ' &=& - (1+w_X) \left( \theta_X + \frac{h '}{2} \right)
+ 3 {\cal H} (w_X - c_X^2) \delta_X \nonumber \\
\theta_X ' &=& - {\cal H} ( 1 - 3 c_X^2) \theta_X + \frac{c_X^2}{1+w_X}
k^2 \delta_X \,,
\label{perturbations}
\end{eqnarray}
where $h$ is the trace of the metric perturbations in the synchronous
gauge \cite{mabert}. This set of equations agrees with those used in
\cite{fabris}. In order to study the Jeans instability for a GCG it
is useful to study the equation for the (gauge invariant) comoving density
contrast $\Delta_X = \delta_X + 3 (1 + w_X) {\cal H} \theta_X/k^2$ in the
approximation in which the GCG is the only component of the universe:
\begin{eqnarray}
\Delta_X '' &+& {\cal H} \left( 1 + 3 c_X^2 - 6 w_X \right) \Delta_X '
\nonumber \\
&+& c_X^2 k^2 \Delta_X - \frac{3}{2} {\cal H}^2 \left( 1 + 8 w_X - 3 w_X^2
- 6 c_X^2  \right) \Delta_X = 0 \,.
\label{jeans}
\end{eqnarray}
We see that the Jeans instability of a GCG is very similar to CDM in the
dust limit (when $w_X \sim c_X^2 \sim 0$). 
However, because of the time dependence of $w_X$ the Jeans instability is
progressively removed since the quantity $1 + 8 w_X - 3 w_X^2
- 6 c_X^2$ changes sign as $w_X$ departs from $0$.
At the same time, GCG perturbations start to oscillate as soon as $c_X^2$
draws away from $0$. We confirm numerically this behaviour, also when
other fluids are present. In Fig. (\ref{figcontrasti}) we show the
comparison of the evolution of cosmological perturbations in GCG models
(without CDM) and $\Lambda$CDM models.

We have implemented Eqs. (\ref{solution}) and (\ref{perturbations}) in a
modified version of CMBFAST \cite{sz}. We have tested the code against 
the sCDM model obtained by setting $A=0$ in Eq. (\ref{solution}) and we
have considered an initial adiabatic scale invariant spectrum for 
perturbations. Instead, by setting $B=0$ in Eq. (\ref{solution}) one
obtains a $\Lambda$CDM model in the background, but with not-trivial
perturbations in the dark energy sector (see Eq. (\ref{jeans})). 

\begin{figure}[!t]
\begin{tabular}{cc}
\epsfxsize=1.7in \epsfbox{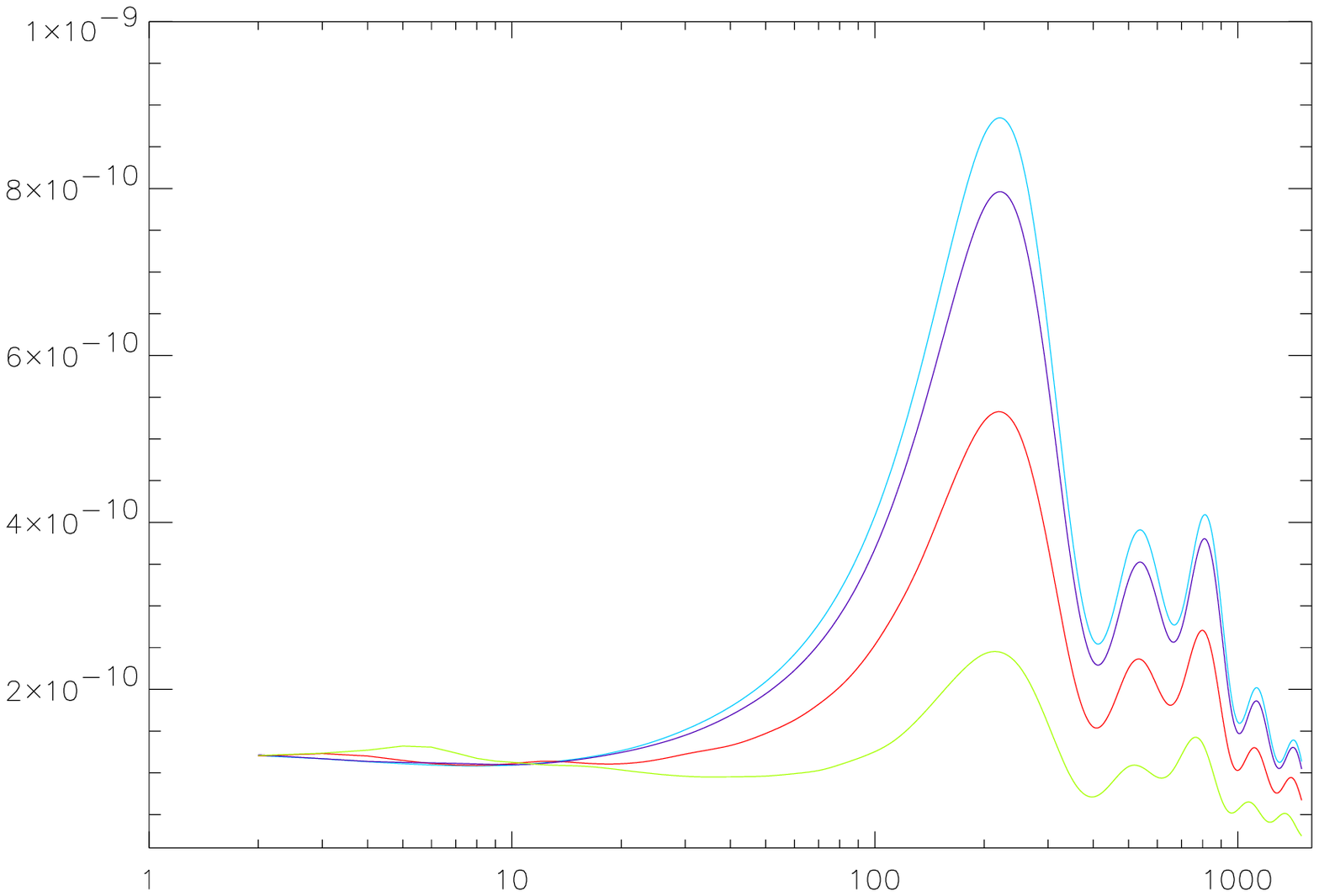} &
\epsfxsize=1.7in \epsfbox{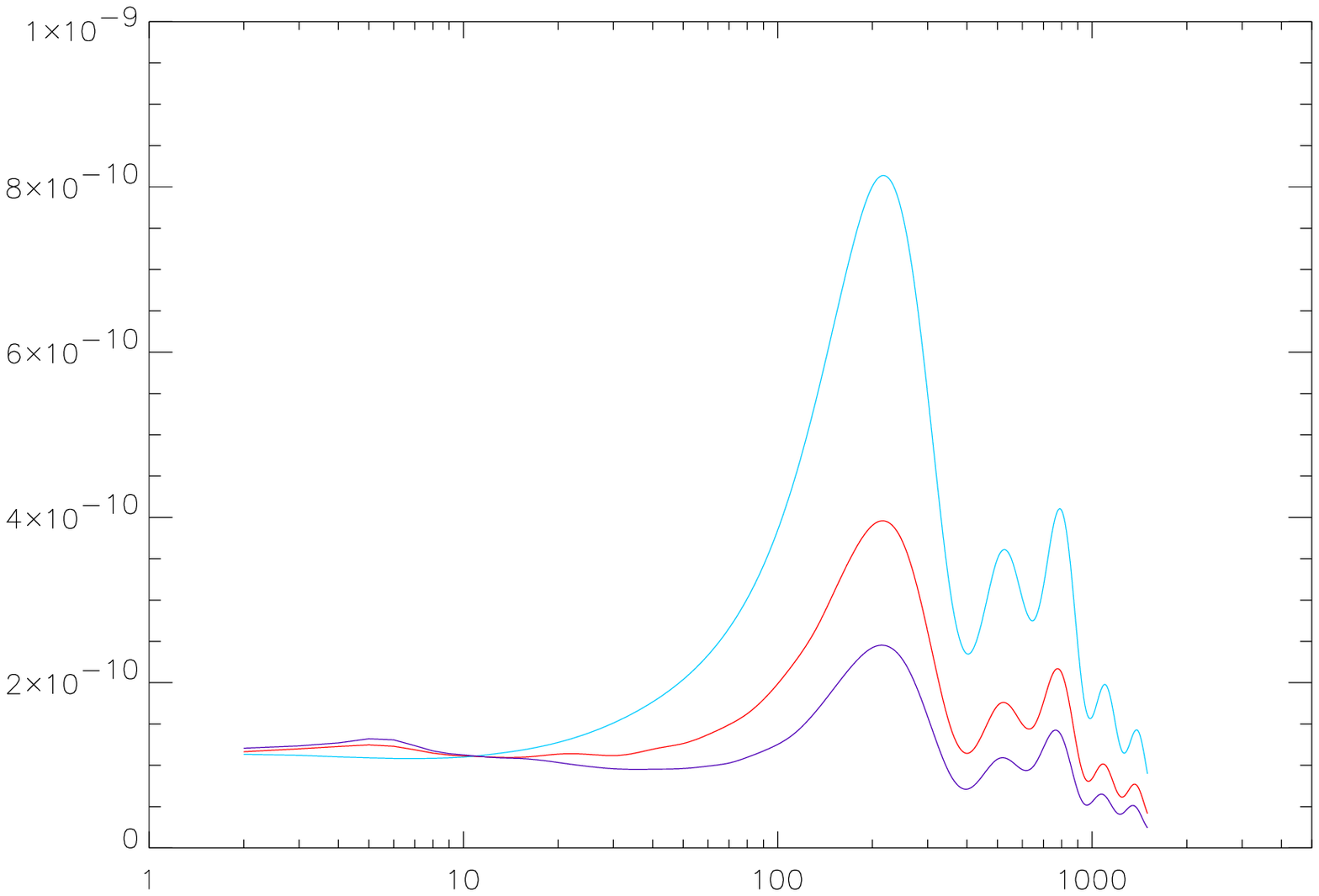}
\end{tabular}
\caption{$C_\ell$ spectra of temperature anisotropies $\delta T/T$ versus 
$\ell$ for CGC varying the ratio $B/A$ (on the left) and $\alpha$ (on the
right). The parameters are the same of Fig. 1 except $B/A = 0.1 \,, 
0.01 \,, 0.001$ and $\Lambda$CDM (from bottom to up, on the left) and
$\alpha = 1, 0.5, 0$ (from bottom to up, on the right) for $B/A= 0.1$.}
\label{doppio}
\end{figure}

In Fig. (\ref{doppio}) we show the dependence of the spectrum of
temperature anisotropies on the ratio $B/A$ and $\alpha$. The three
parameters $A, B, \alpha$ are in direct relation with the physical
quantities of dark energy at present time, as $\Omega_{X 0}, w_{X 0}, \dot
w_{X 0}$. Therefore these parameters can be constrained by
maximizing the likelihood function with the present data \cite{new}. In
particular, from the left panel of Fig. (\ref{doppio}) one can see how 
spectra can be sensibly different for a GCG model which differs from a
$\Lambda$CDM model by less than $10 \%$ in the background evolution.

\begin{figure}
\epsfxsize=3. in \epsfbox{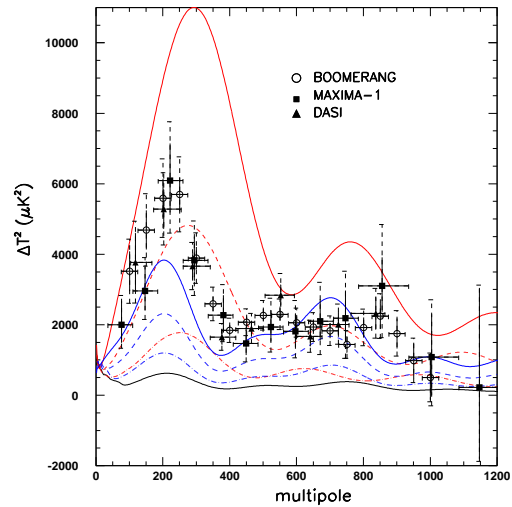}
\caption{Spectra of temperature anisotropies $\delta T$ for a unified
model ($\Omega_{X0}=.95$
and $h=70$) varying $B/A$ compared with the
BOOMERANG, MAXIMA, DASI data. The solid line at the bottom is $B/A=1$.
The blue lines correspond to $10, 50$ and a sCDM respectively from bottom
to top. The red lines correspond to $.01, .001$ and a
$\Lambda$-baryonmodel from bottom to up.} 
\label{figdatabsua}
\end{figure}

\begin{figure}
\epsfxsize=3.in \epsfbox{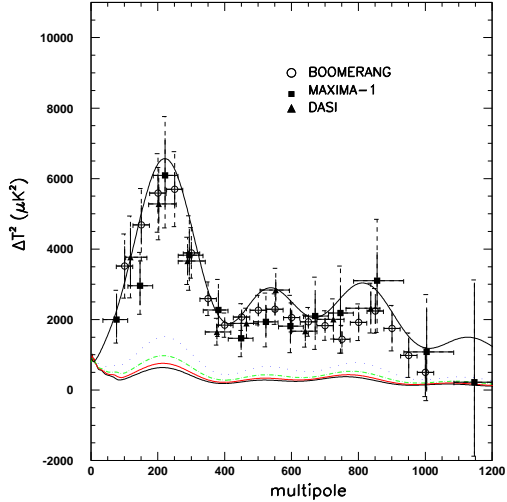}
\caption{Spectra of temperature anisotropies $\delta T$ for a unified
model ($\Omega_{X0}=.95$, $h=70$, $B/A=.36$) varying $\alpha$ ($1, 0.75,
0.5, 0.25, 0$, from bottom to top) compared with the BOOMERANG, MAXIMA,
DASI data.}
\label{figdataalfa}
\end{figure}

Because of its early dust behaviour, a GCG may also represent a
prototypical unified model of dark matter and dark energy (see 
\cite{padcho} for a similar proposal, but with a scale dependent
equation of state). In Figs. (\ref{figdatabsua},\ref{figdataalfa}) we
compare GCG models
without CDM ($h=.7 \,, \Omega_{X 0}=.95$) with the BOOMERANG
\cite{boom}, MAXIMA \cite{maxi} and DASI \cite{dasi} data, when varying
$B/A$ and $\alpha$. In particular, in Fig (\ref{figdatabsua}) we see how
all the models lie below the limiting case of sCDM ($A=0$) and a
$\Lambda$-baryon model (very similar, but {\em not} equivalent to $B=0$).
The resulting spectrum of CMB anisotropies constrain these GCG models
\cite{new} much more than the SN Ia data \cite{SNconstraint}.

{\bf 4.} We have studied the implications on the
evolutions of cosmological perturbations and on CMB anisotropies 
of a GCG as a candidate for dark
energy. This GCG covers all the interesting possible cases of a dark
energy model from a polytropic gas. 
A GCG is more distinguishable from a $\Lambda$CDM model than a
QCDM model \cite{lversusq} since both GCG background and perturbations are
important not only at late redshift, when the GCG behaves like a
cosmological constant. Infact the GCG plays a role of dust component
before turning in a cosmological constant, modifying not only the
positions of the peaks, but the overall shape of the $C_\ell$ because 
of a big Integrate Sachs-Wolfe (ISW) effect. In this sense, the QCDM
models with standard scalar fields \cite{rp,accellera} are the
most economic way to modify a $\Lambda$CDM model in a dark energy model.
The dark component sector(s) may be much more obscure and less simple
\cite{hu}, and the GCG models are an example of this. The next
CMB experiments \cite{map,planck} and LSS data will be be very hepful in
constraining the physical properties of the dark component sector(s).

\vspace{.2cm}

{\bf Acknowledgments}

\noindent
We would like to thank L. Popa and M. Sandri for useful suggestions.

\vspace{.2cm}

{\bf Note added:} While this paper was in preparation a preprint
\cite{gcgpeaks} by Bento et al., which studies the location of the CMB
peaks in presence of a unified GCG model within an
analytic approximation, appeared. We have checked their analytic results
with our code and we have found a systematic overestimation of the
peak positions (in particular for the third peak).

\end{document}